\documentclass[aps,prb,twocolumn,superscriptaddress,showpacs,floatfix,amsmath,amssymb]{revtex4}
\usepackage{graphicx}% Include figure files
\usepackage{graphicx}% Include figure files
\usepackage{dcolumn}% Align table columns on decimal point
\usepackage{bm}% bold math
\usepackage[below]{placeins}

\begin{document}

%\preprint{}

\title{Magnetic Order in TbCo$_{2}$Zn$_{20}$ and TbFe$_{2}$Zn$_{20}$}

% Please use the appropriate macro for each type of information

%\author{}
%\thanks{}
%\affiliation{}

\author{W. Tian}
\affiliation{Ames Laboratory and Department of Physics and
Astronomy, Iowa State University, Ames, Iowa 50011, USA}

\author{A. D. Christianson}
\affiliation{Oak Ridge National Laboratory, Oak Ridge, TN 37831,
USA}

\author{J. L. Zarestky}
\affiliation{Ames Laboratory and Department of Physics and
Astronomy, Iowa State University, Ames, Iowa 50011, USA}

\author{S. Jia}
 \altaffiliation{Present address: Department of chemistry, Princeton
University, Princeton, NJ 08544, USA} \affiliation{Ames Laboratory
and Department of Physics and Astronomy, Iowa State University,
Ames, Iowa 50011, USA}
\author{S. L. Bud'ko}
\author{P. C. Canfield}
\affiliation{Ames Laboratory and Department of Physics and
Astronomy, Iowa State University, Ames, Iowa 50011, USA}

\author{P. M. B. Piccoli}

\author{A. J. Schultz}
\affiliation{Argonne National Laboratory, Argonne, IL 60439, USA}
% add more authors

\begin{abstract}

We report neutron diffraction studies of TbCo$_2$Zn$_{20}$ and
TbFe$_2$Zn$_{20}$, two isostructural compounds which exhibit
dramatically different magnetic behavior. In the case of
TbCo$_2$Zn$_{20}$, magnetic Bragg peaks corresponding to
antiferromagnetic order are observed below $T_N$ $\approx$ 2.5 K
with a propagation vector of (0.5 0.5 0.5). On the other hand,
TbFe$_2$Zn$_{20}$ undergoes a ferromagnetic transition at
temperatures as high as 66 K which shows a high sensitivity to
sample-to-sample variations. Two samples of TbFe$_2$Zn$_{20}$ with
the same nominal compositions but with substantially different
magnetic ordering temperatures ($T_c$ $\approx$ 51 and 66 K) were
measured by single crystal neutron diffraction. Structural
refinements of the neutron diffraction data find no direct signature
of atomic site disorder between the two TbFe$_2$Zn$_{20}$ samples
except for subtle differences in the anisotropic thermal parameters.
The differences in the anisotropic thermal parameters between the
two samples is likely due to very small amounts of disorder. This
provides further evidence for the extreme sensitivity of the
magnetic properties of TbFe$_2$Zn$_{20}$ to small sample variations,
even small amounts of disorder.
\end{abstract}

\pacs{valid numbers to be inserted here
      }

\maketitle

\section{Introduction}

The RT$_2$Zn$_{20}$ (R = Rare Earth and T = Transition metal) family
of materials exhibit diverse magnetic properties
\cite{nature-physics-2007,Torikachvili-2006,Jia-2007, Jia-2008,
Jia-2009}. For example, nearly ferromagnetic Fermi liquid behaviour
was observed in YFe$_2$Zn$_{20}$ \cite{nature-physics-2007} and
LuFe$_2$Zn$_{20}$ \cite{Jia-2009}. Subsequently it was shown that by
substituting a small amount of Y by Gd the
Gd$_x$Y$_{1-x}$Fe$_2$Zn$_{20}$ system becomes ferromagnetic for $x$
$>$ 0.02 \cite{Jia-2007} and a remarkably high ferromagnetic
ordering temperature of $T_c$ = 86 K is found for GdFe$_2$Zn$_{20}$
\cite{nature-physics-2007}. Moreover, six closely related Yb-based
heavy fermion compounds were discovered in YbT$_2$Zn$_{20}$ for T =
Fe, Ru, Os, Co, Rh and Ir \cite{Torikachvili-2006} that
significantly increase the total number of Yb-based heavy fermion
materials. The exotic magnetic properties discovered in recent
studies have generated great interest in the RT$_2$Zn$_{20}$
compounds.

\begin{figure}
\centering\includegraphics[width=3.4in]{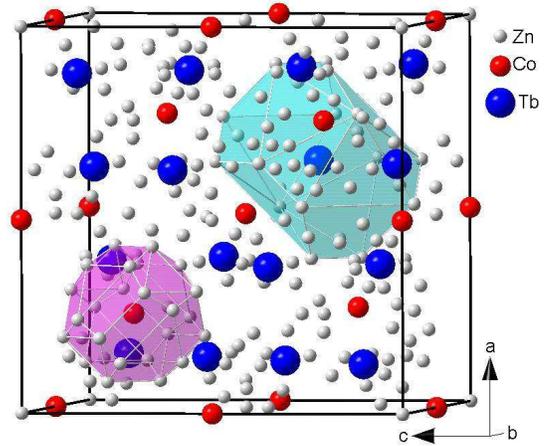}
\caption{\label{fig:crystal-structure}(Color online) The schematic
crystal structure of TbCo$_{2}$Zn$_{20}$ shows that both Tb and Co
ions are surrounded exclusively by Zn atoms as illustrated by the
coordination polyhedron of Tb (cyan) and Co (pink).}
\end{figure}

The RT$_2$Zn$_{20}$ materials crystallize in the cubic
CeCr$_2$Al$_{20}$ type structure ($\textit{Fd}$$\bar{3}$$\textit{m}$
space group). The R and T ions each occupy an unique
crystallographic site, 8$\textit{a}$ and 16$\textit{d}$
respectively, whereas the Zn ions have three distinct
crystallographic sites, 96$\textit{g}$, 48$\textit{f}$ and
16$\textit{c}$ \cite{Verena-1998,Moze-1998}. Even though these
materials are comprised of more than 85$\%$ Zn, they are highly
tunable and ideal for fundamental research such as magnetism for two
reasons. (1) As illustrated in Fig.~\ref{fig:crystal-structure}, the
R and T ions are surrounded exclusively by Zn atoms and there are no
R-R, R-T, and T-T nearest neighbors in the system; (2) Due to the
very low R concentrations in these compounds (less than 5 at.$\%$),
the local environment of the R site is only weakly affected by the
changing of T. Thus by substituting on either the rare-earth site or
the transition metal site, these isostructural compounds can serve
as model systems for a variety of physical phenomenon.

%Figure 1 TbCo2Zn20 crystal structure, Tb cage.

Among the RT$_2$Zn$_{20}$ compounds, GdFe$_2$Zn$_{20}$ and
GdCo$_2$Zn$_{20}$ have attracted special attention due to the
distinct magnetic properties exhibited by each
material\cite{nature-physics-2007}. GdFe$_2$Zn$_{20}$ is a
ferromagnet with a remarkably high ordering temperature of $T_c$ =
86 K. However, upon substitution of Co for Fe, ferromagnetism is
rapidly suppressed\cite{Jia-2008} culminating in antiferromagnetic
order at $T_N$= 5.7 K in GdCo$_2$Zn$_{20}$. Band structure
calculations suggest that the enhanced magnetic ordering temperature
in GdFe$_2$Zn$_{20}$ is due to a large d-electron contribution to
the electronic density of states when compared to GdCo$_2$Zn$_{20}$.
In addition to sensitivity to transition metal substitution,
GdFe$_2$Zn$_{20}$ exhibits sensitivity to small sample-to-sample
variations. Similar behavior has also been observed in Sc$_3$In and
ZrZn$_2$, materials on the ferromagnetic side just over the Stoner
limit \cite{Moriya-1985,Brommer-1990,Zellermann-2004}.

To provide further information concerning the microscopic origin of
the magnetic behavior in the RT$_2$Zn$_{20}$ family we have
performed neutron scattering measurements on TbFe$_2$Zn$_{20}$ and
TbCo$_2$Zn$_{20}$.  TbFe$_2$Zn$_{20}$ and TbCo$_2$Zn$_{20}$ are
ideal for neutron scattering studies as these materials exhibit
similar magnetic behavior to GdFe$_2$Zn$_{20}$ and
GdCo$_2$Zn$_{20}$, but do not possess the strong neutron absorption
cross-section of gadolinium. We have studied single crystalline and
polycrystalline samples of TbCo$_2$Zn$_{20}$ and find
antiferromagnetic order below $T_N$ $\approx$ 2.5 K with a
propagation vector of (0.5 0.5 0.5). Refinements of the magnetic
structure indicate an ordered moment of $\sim$7.46 $\mu_B$ at 1.4 K.
In the case of ferromagnetic TbFe$_2$Zn$_{20}$ we investigated two
single crystals prepared by slightly different methods that resulted
in a $\sim$15 K difference in ordering temperatures.  No obvious
signs of site disorder were detected in structural refinements of
the neutron data, highlighting the extreme sensitivity to small
sample variations. However, there are small differences in the
refinements which suggest that the sample with the higher transition
temperature is more highly ordered. The neutron scattering data are
consistent with a picture of the RFe$_2$Zn$_{20}$ family where the
high rare earth ordering temperatures are associated with the highly
polarizable Fe lattice and are extremely sensitive to disorder.

\section{Experimental Details}

TbFe$_2$Zn$_{20}$ and TbCo$_2$Zn$_{20}$ single crystals were grown
from a Zn-rich self flux \cite{nature-physics-2007, Canfield-1992}
in the novel materials and ground states group at Ames Laboratory.
Two methods, denoted method 1 and 2 were used to grow crystals of
TbFe$_2$Zn$_{20}$: s1 and s2, respectively. The primary difference
between methods 1 and 2 is that the initial molar ratios of the
starting elements (Tb:Fe:Zn) were 2:5:95 and 2:3:93 respectively.
The resulting single crystals have identical morphology and size to
single crystals grown from the molar ratio of the starting elements
R:T:Zn = 2:4:94 in Ref.~\onlinecite{nature-physics-2007}. Although
the magnetic ordering temperatures of the TbFe$_2$Zn$_{20}$ samples
synthesized by methods 1 and 2 are different by $\sim$ 15 K (see
Fig.~\ref{fig:FeBulk} below), single crystal x-ray diffraction
measurements cannot fully resolve possible variations of occupancy
of Fe and Zn sites, due to their similar atomic
numbers.\cite{tfexray}  The TbCo$_2$Zn$_{20}$ crystals were grown
from the molar ratio of the starting elements Tb:Co:Zn = 2:4:94. The
magnetic susceptibility of TbCo$_2$Zn$_{20}$ and TbFe$_2$Zn$_{20}$
were measured using a Quantum Design MPMS units whereas transport
measurements on TbFe$_2$Zn$_{20}$ were made using Quantum Design
PPMS.

\begin{table*}[htp!]
\caption{\label{tb:CoRef} Rietveld structure refinement results for
TbCo$_2$Zn$_{20}$ at T = 1.4 K and 10 K.}
\begin{ruledtabular}
\begin{tabular}{p{62pt}ccccc}
\centering{Atom} & site & T(K) & $x$ & $y$ & $z$\\
\hline
\centering {Tb} & 8$a$ & 10 & 0.125& 0.125 & 0.125\\
\centering {Co} & 16$d$ & 10 & 0.5 & 0.5 & 0.5 \\
\centering {Zn1} &16$c$ & 10 & 0 & 0 & 0 \\
\centering {Zn2} & 48$f$ & 10 & 0.4916(6) & 0.125 & 0.125\\
\centering {Zn3} & 96$g$ & 10 & 0.0608(3) & 0.0608 & 0.3245(5) \\
\centering{Lattice\par and\par Reliability\par factors}& & 10 &
\multicolumn{3}{p{230pt}}{\centering{$a$=$b$=$c$=14.0358(3) \AA;
$V_{cell}$=2765.104(8) \AA$^{3}$; R$_{Bragg}$=5.28 \%; R$_{p}$=5.0
\%; R$_{wp}$=6.4 \%; $\chi
^{2}$=1.4;}}\\
 & & 1.4 & \multicolumn{3}{p{240pt}}{\centering{$a$=$b$=$c$=14.0290(2) \AA; $V_{cell}$=2761.09(8) \AA$^{3}$;
 R$_{Bragg}$=8.35\%; \\R$_{mag}$=13.86 \%; R$_{p}$=13.8 \%;
R$_{wp}$=15.5\%; $\chi ^{2}$=12.8;}}\\
\end{tabular}
\end{ruledtabular}
\end{table*}

Neutron diffraction measurements of TbCo$_2$Zn$_{20}$ were performed
on a $\sim$ 1.7 gram single crystal and $\sim$ 2 gram powder using
the HB1A triple axis spectrometer at the High Flux Isotope Reactor
(HFIR) at the Oak Ridge National Laboratory (ORNL). The HB1A
spectrometer operates with a fixed incident energy of 14.64 meV
using a double pyrolitic graphite (PG) monochromator system. The
second-order contamination in the beam was removed
($\textit{I}_{\lambda/2}$ $\cong$ 10$^{-4}$$\textit{I}_{\lambda}$)
by placing two PG filters located before and after the second
monochromator. The TbCo$_2$Zn$_{20}$ single crystal was mounted on a
aluminum plate and oriented in the ($\textit{hhl}$) scattering
plane.  For low temperature measurements, both samples were sealed
in an aluminum can under Helium atmosphere and cooled using a
closed-cycle Helium refrigerator. A collimation of
48'-48'-sample-40'-68' from reactor to detector was used throughout
the measurements and all results shown have been normalized to a
beam monitor count.

%Figure 2 Single crystal (hhl) long scan LT&HT, order parameter.
\begin{figure}
\centering\includegraphics[width=3.4in]{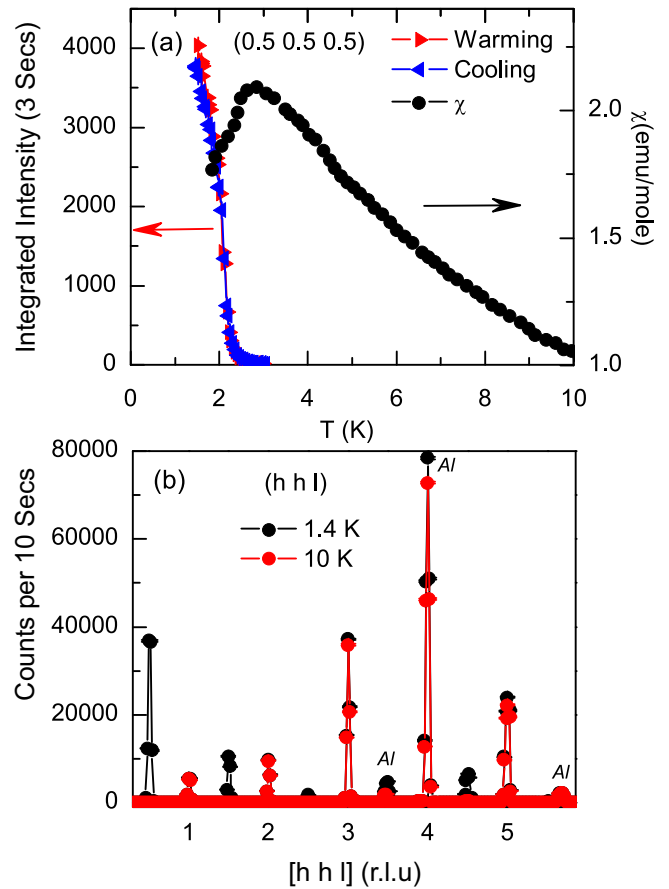}
\caption{\label{fig:crystal}(Color online) TbCo$_2$Zn$_{20}$ single
crystal measurements with the crystal oriented in the
($\textit{hhl}$) scattering plane. (a) Temperature dependence of the
integrated scattering intensity of (0.5 0.5 0.5) magnetic reflection
(left) and the magnetic susceptibility (right) measured with $H
\parallel [111]$ and $H = 0.1$ T. Note the temperature of onset magnetic
scattering is close to the temperature (2.5 K) of maximal
$d$($\chi$$T$)/$d$$T$ value in Ref. \onlinecite{Jia-2009}. (b)
Typical elastic long scans along the ($\textit{hhl}$) direction
measured at T = 1.4 K and T = 10 K, respectively. Aluminum
reflection contributions (2$\theta$ = 60.79$^\circ$, 71.5$^\circ$,
and 111.44$^\circ$) from the aluminum sample holder are marked.}
\end{figure}

Two TbFe$_2$Zn$_{20}$ single crystals with the same nominal
compositions hereafter denoted s1 and s2 were selected for neutron
scattering measurements. The crystals were cut to
$\sim$3$\times$3$\times$3 mm$^3$ with sample masses of 0.213 g (s1)
and 0.210 g (s2). The samples were mounted on aluminum pins with
epoxy and attached to the cold finger of a standard closed cycle
refrigerator. Neutron diffraction patterns were collected in a
standard configuration of the SCD single-crystal neutron Laue
Diffractometer at the Intense Pulsed Neutron Source (IPNS) at
Argonne National Laboratory\cite{SCD}. 10 crystal settings were
selected to cover an octant of reciprocal space for each sample
resulting in around 3600 reflections per sample. A counting time of
3 hours per crystal setting was used. Data were reduced with the
ISAW package \cite{ISAW} and corrected for neutron absorption.
Structural refinements were done with the General Structural
Analysis System (GSAS) software package\cite{gsas}. Equivalent
reflections have not been averaged since the extinction correction
applied during the structural refinements is strongly wavelength
dependent.

%Figure 3 Powder Diffraction refinements LT&HT.
\begin{figure}
\centering\includegraphics[width=3.4in]{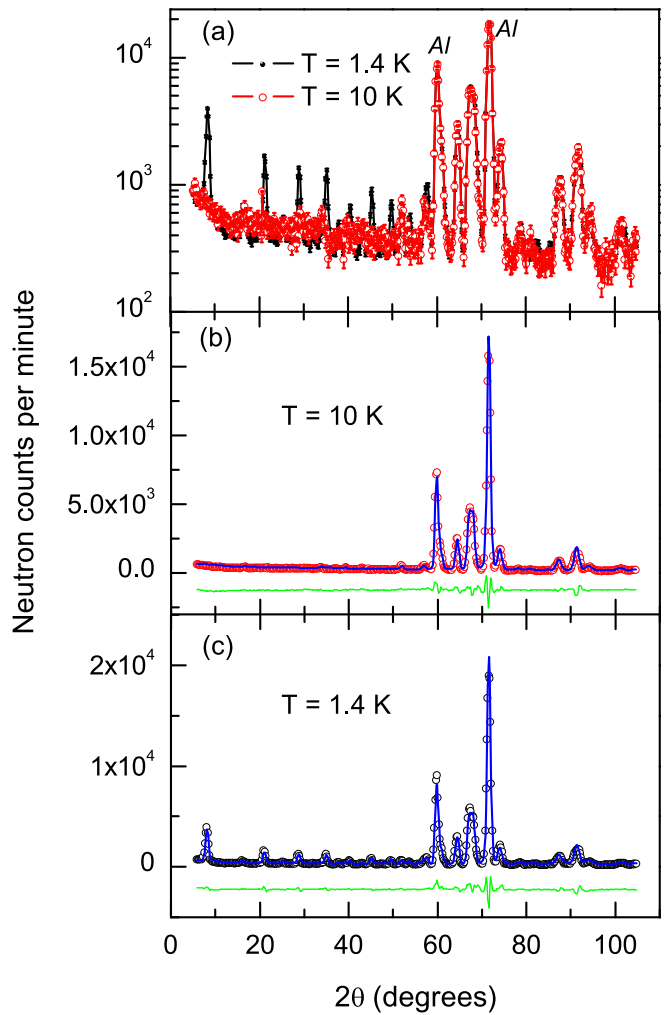}
\caption{\label{fig:powderLTvsHT}(Color online) (a) Neutron powder
diffraction patterns in log scale measured at $T$ = 1.4 K (solid
circles) and $T$ = 10 K (open circles). Aluminum reflections
(2$\theta$ = 60.79$^\circ$, 71.5$^\circ$) from aluminum sample
container are marked. Observed neutron powder diffraction pattern
(circles), calculated profile (solid line) and their difference
(bottom line) between $5 < 2\theta < 105$ measured at (b) $T=$ 10 K,
and (c) $T=$ 1.4 K.}
\end{figure}

\section{Results}

\subsection{TbCo$_2$Zn$_{20}$}

Figure 2(a) shows the magnetic susceptibility (right axis) of
TbCo$_2$Zn$_{20}$ measured at $H$ = 0.1 T with $H
\parallel [111]$.  A peak observed at $T \approx 2.8$ K with a
maximal value of $d$($\chi$$T$)/$d$$T$ found for $T \sim$ 2.5 K
\cite{Jia-2009} suggests a magnetic, long-range order (LRO)
transition near 2.5 K. To examine this low temperature phase, single
crystal neutron diffraction measurements were carried out at
temperatures above and below $T \approx 2.5$ K.
Fig.~\ref{fig:crystal} (b) compares two wave vector scans measured
along the ($\textit{hhl}$) direction at T = 1.4 K and 10 K. At T =
1.4 K, $(\frac{h}{2}, \frac{h}{2},\frac{l}{2})$ ( $h$, $l$ are odd
integers) type reflections were detected. The absence of these
reflections at T = 10 K indicate they are magnetic in origin. To
verify the observed magnetic reflections at 1.4 K are indeed
associated with the $T \approx 2.5$ K phase transition, order
parameter measurements were carried out by monitoring the strong
(0.5 0.5 0.5) magnetic peak as a function of temperature between $T
= 1.4$ K and $T = 10$ K. As shown in Fig.~\ref{fig:crystal} (a), the
integrated intensity of (0.5 0.5 0.5) (right axis) increases rapidly
at the transition temperature $T_N \approx 2.5$ K in good agreement
with the magnetic susceptibility \cite{Jia-2009}. The single crystal
measurements confirm that TbCo$_2$Zn$_{20}$ is a long-range ordered
antiferromagnet.

To study the low temperature magnetic structure of
TbCo$_2$Zn$_{20}$, $\sim$ 2 grams powder were prepared by grinding
single crystals. Powder diffraction data were collected over the
range of 5$^\circ < 2\theta < 165^\circ$ with a $2\theta$ step size
of 0.02$^\circ$ at 1.4 K $\&$ 10 K. As illustrated in
Fig.~\ref{fig:powderLTvsHT} (a), multiple magnetic peaks were
observed below T$_N$ associated with the antiferromagnetic long
range ordered phase. Consistent with the single crystal
measurements, the observed magnetic peaks can be indexed as
$(\frac{h}{2}, \frac{k}{2},\frac{l}{2})$ ($h$, $k$, $l$ are odd
integers) suggesting the doubling of the magnetic unit cell along
all three principle axis directions.

We carried out Rietveld refinements of the neutron data using the
FULLPROF program \cite{Carvajal-1993}. The T = 10 K neutron powder
data was first refined to check the nuclear crystal structure of
TbCo$_2$Zn$_{20}$. The refinement results show
(Fig.~\ref{fig:powderLTvsHT} (b)) that the crystal structure of
TbCo$_2$Zn$_{20}$ belongs to the cubic
$\textit{Fd}$$\bar{3}$$\textit{m}$ space group consistent with
earlier studies. The refined lattice constants and atomic positions
are given in Table~\ref{tb:CoRef}. To refine the T = 1.4 K low
temperature data, the representation analysis technique
\cite{representation-tech}, program SARAH \cite{sarah} and BasIreps
(Rodriguez-Carvajal, 2004) were used as a tool to generate magnetic
structures that are compatible with the TbCo$_2$Zn$_{20}$ crystal
structure. There are four possible irreducible representations
associated with the $\textit{Fd}$$\bar{3}$$\textit{m}$ space group
as listed in table \ref{tb:basis-vector}. The labeling of the
propagation vector and the irreducible representations in table
\ref{tb:basis-vector} follows the scheme used by Kovalev
\cite{Kovalev-1993}. Refinements of the T = 1.4 K data using
FULLPROF indicates that the magnetic structure of TbCo$_2$Zn$_{20}$
can only be described by the irreducible representation $\Gamma_5$.
Figure 3(c) shows the refinement results considering both the
magnetic and crystal structure unit cell. At T = 1.4 K, the refined
magnetic moment of Tb$^{3+}$ ion is ~$\sim$ 7.46 $\mu_{B}$. We note
that the moment has not yet saturated by 1.4 K thus the ordered
moment at T = 0 K is probably somewhat closer to the full moment
value of K $\sim$ 9.5 $\mu_{B}$. Some reduction in the ordered
moment is expected due to crystal field level splitting. Preliminary
inelastic neutron scattering work suggests the crystal field level
splitting of the Tb ground state multiplet in TbCo$_2$Zn$_{20}$ is
less than 1.5 meV.

\begin{table}[htp!]
\caption{\label{tb:basis-vector} Basis vectors (BV) for the space group
$\textit{Fd}$$\bar{3}$$\textit{m}$:2 with ${\bf k}_{-55}=( .5,~ .5,~
.5)$. The Tb atoms of the nonprimitive basis are defined according
to 1: $( .125,~ .125,~ .125)$, 2: $( .875,~ .875,~ .875)$.}
\begin{ruledtabular}
\begin{tabular}{ccc|cccccc}
  IR  &  BV  &  Atom & \multicolumn{6}{c}{BV components}\\
      &      &             &$m_{\|a}$ & $m_{\|b}$ & $m_{\|c}$ &$im_{\|a}$ & $im_{\|b}$ & $im_{\|c}$ \\
\hline
$\Gamma_{2}$ & $\phi_{1}$ &      1 &      1 &      1 &      1 &      0 &      0 &      0  \\
             &              &      2 &      1 &      1 &      1 &      0 &      0 &      0  \\
$\Gamma_{3}$ & $\phi_{2}$ &      1 &      1 &      1 &      1 &      0 &      0 &      0  \\
             &              &      2 &     -1 &     -1 &     -1 &      0 &      0 &      0  \\
$\Gamma_{5}$ & $\phi_{3}$ &      1 &      1 &     -0.5 &     -0.5 &      0 &      0 &      0  \\
             &              &      2 &     -1 &      0.5 &      0.5 &      0 &      0 &      0  \\
             & $\phi_{4}$ &      1 &      0 &  0.866 & -0.866 &      0 &      0 &      0  \\
             &              &      2 &      0 & -0.866 &  0.866 &      0 &      0 &      0  \\
$\Gamma_{6}$ & $\phi_{5}$ &      1 &      1 &     -0.5 &     -0.5 &      0 &      0 &      0  \\
             &              &      2 &      1 &     -0.5 &     -0.5 &      0 &      0 &      0  \\
             & $\phi_{6}$ &      1 &      0 &  0.866 & -0.866 &      0 &      0 &      0  \\
             &              &      2 &      0 &  0.866 & -0.866 &      0 &      0 &      0  \\
\end{tabular}
\end{ruledtabular}
\end{table}

\subsection{TbFe$_2$Zn$_{20}$}

\begin{table*}[htp!]
\caption{\label{tb:Feref1} Single-crystal structural refinement
results for TbFe$_2$Zn$_{20}$}
\begin{ruledtabular}
\begin{tabular}{p{62pt}ccccc}
\centering{Atom} & S\# & $x$ & $y$ & $z$ & fraction \\
\hline
\centering {Tb (8$a$)}& s1 100 K & 0.125& 0.125 & 0.125 & 1\\
 & s2 100 K & 0.125 & 0.125 & 0.125 & 1\\
 & s2 20 K & 0.125 & 0.125 & 0.125 & 1\\
\centering {Fe (16$d$)}& s1 100 K & 0.5 & 0.5 & 0.5 & 1.006(3)\\
& s2 100 K & 0.5 & 0.5 & 0.5 & 0.986(4)\\
& s2 20 K & 0.5 & 0.5 & 0.5 & 0.986\\
\centering {Zn1 (16$c$)}& s1 100 K & 0 & 0 & 0 & 0.976(4)\\
& s2 100 K & 0 & 0 & 0 & 0.993(5)\\
& s2 20 K & 0 & 0 & 0 & 0.993\\
\centering {Zn2 (48$f$)}& s1 100 K & 0.48932(2) & 0.125 & 0.125 & 0.993(3) \\
& s2 100 K & 0.48923(2) & 0.125 & 0.125 & 0.979(4)\\
& s2 20 K & 0.48920(4) & 0.125 & 0.125 & 0.979\\
\centering {Zn3 (96$g$)}& s1 100 K & 0.05884(1) & 0.05884 & 0.32620(1) & 0.991(3) \\
& s2 100 K & 0.05886(1) & 0.05886 & 0.32612(2) & 0.986(3)\\
& s2 20 K & 0.05885(2) & 0.05885 & 0.32611(2) & 0.986\\
\centering{Lattice cell/\par Reliability\par factors} & s1 100 K &
\multicolumn{3}{p{195pt}}{\centering{$a$=14.049(2) \AA;  $V_{cell}$=2772.9(7) \AA$^{3}$ \\
R=5.7 \%; R$_{w}$=5.3
\%; $\chi ^{2}$=2.7;}}\\
 & s2 100 K & \multicolumn{3}{p{195pt}}{\centering{$a$=14.049(3) \AA; $V_{cell}$=2772.9(9) \AA$^{3}$;
R=6.4 \%;
R$_{w}$=6.3\%; $\chi ^{2}$=2.1 ;}}\\
& s2 20 K & \multicolumn{3}{p{195pt}}{\centering{$a$=14.038(3) \AA;
$V_{cell}$=2766.4(9) \AA$^{3}$;
 R=6.0\%;
R$_{w}$=6.7\%; $\chi ^{2}$=2.4; $M_{20 K}=6.01(5)\mu_B$}}\\
\end{tabular}
\end{ruledtabular}
\end{table*}

\begin{table*}[htp!]
\caption{\label{tb:Feref2} Anisotropic Thermal Parameters for
TbFe$_2$Zn$_{20}$}
\begin{ruledtabular}
\begin{tabular}{p{62pt}ccccccc}
\centering{Atom} & s\# & 100*U11 & 100*U22 & 100*U33 & 100*U12 & 100*U13 & 100*U23\\
\hline
\centering {Tb (8$a$)}& S1 100 K & 0.297(11) & 0.297 & 0.297 & 0 & 0 & 0  \\
 & S2 100 K & 0.329(12) & 0.329 & 0.329 & 0 & 0 & 0\\
\centering {Fe (16$d$)}& S1 100 K & 0.345(6) & 0.345 & 0.345 & -0.011(5) & -0.011 & -0.011 \\
& S2 100 K & 0.370(7) & 0.370 & 0.370 & -0.024(5) & -0.024 & -0.024\\
\centering {Zn1 (16$c$)}& S1 100 K & 0.715(12) & 0.715 & 0.715 &
-0.135(9) & -0.135 & -0.135
\\
& S2 100 K & 0.807(13) & 0.807 & 0.807 & -0.135(9) & -0.135 & -0.135 \\
\centering {Zn2 (48$f$)}& S1 100 K & 0.452(10) & 0.428(7) & 0.428(7) & 0 & 0 & -0.075(7)\\
& S2 100 K & 0.496(11) & 0.452(8) & 0.452(8) & 0 & 0 & -0.080(8) \\
\centering {Zn3 (96$g$)}& S1 100 K & 0.608(5) & 0.608 & 0.464(7) & -0.143(6) & -0.019(4) & -0.019(4)\\
& S2 100 K & 0.661(6) & 0.661 & 0.517(8) & -0.145(6) & -0.020(4) &
-0.020(4)
\\
\end{tabular}
\end{ruledtabular}
\end{table*}

%Figure 4 Bulk property of TbFe$_2$Zn$_{20}$.
\begin{figure}
\centering\includegraphics[width=3.4in]{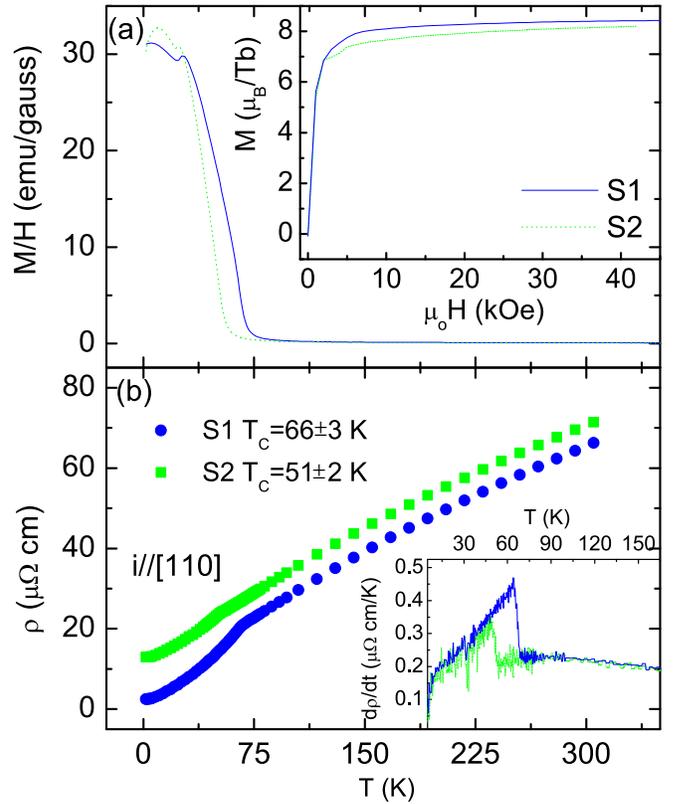}
\caption{\label{fig:FeBulk}(Color online) Physical properties of
TbFe$_2$Zn$_{20}$ crystals, s1 and s2.  (a) M/H as a function of
temperature. Inset: Magnetization versus applied magnetic field H at
T = 2 K with H $\parallel$ [111] (b) Electrical resistivity $\rho$
as a function of temperature. Inset: d$\rho$$/$dt between 1.5 K and
165 K.}
\end{figure}

%Figure 5 SCD reciprocal space TbFe$_2$Zn$_{20}$.
\begin{figure}
\centering\includegraphics[width=3.4in]{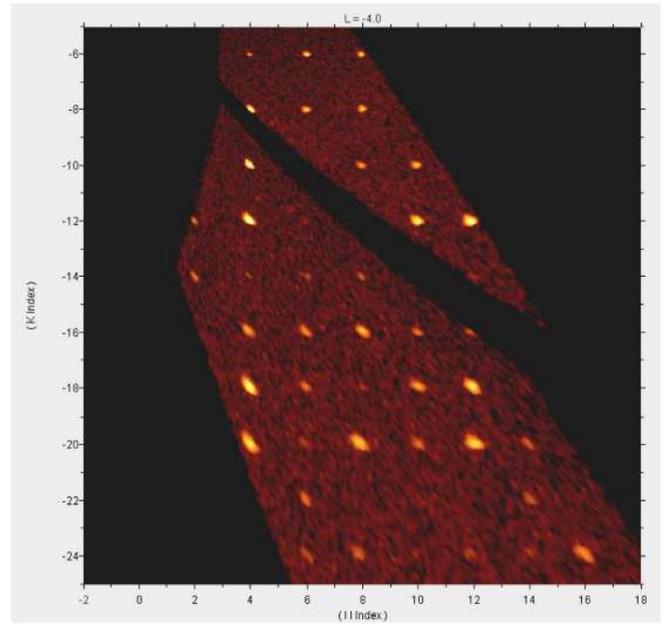}
\caption{\label{fig:scddata}(Color online) Diffraction data for
TbFe$_2$Zn$_{20}$ (s2). Representative data collected at 100 K with
the SCD instrument.}
\end{figure}

As noted above, two single crystals of TbFe$_2$Zn$_{20}$ (s1 and s2)
were studied. Although s1 and s2 have the same nominal stoichiometry
, an $\sim$ 15 K difference in T$_c$ was observed in thermodynamic
and transport properties. The temperature dependence of the
magnetization divided by applied field (M/H) (Fig.~\ref{fig:FeBulk}
(a)) and resistivity (Fig.~\ref{fig:FeBulk} (b)) reveal a FM
transition temperature $T_{c1}$ $\approx$ 66 $\pm$ 3 K for s1 and
$T_{c2}$ $\approx$ 51 $\pm$ 2 K for s2 respectively. From the
resistivity data, the calculated Residual Resistivity Ratio (RRR)
for samples s1 and s2 are RRR(s1) $\approx$ 26.3 and RRR(s2)
$\approx$ 5.5. The Residual Resistivity Ratio is the ratio between
the low temperature electrical conductivity (e.g. at 2 K below the
boiling point of liquid helium) and the room temperature electrical
conductivity,
RRR=$\sigma_{2K}/\sigma_{300K}$=$\rho_{300K}/\rho_{2K}$. Because the
room temperature electrical conductivity is primarily due to phonon
scattering, whereas the low temperature electrical conductivity is
primarily due to disorder scattering, therefore the lower $T_c$
observed for s2 may be associated with the lower RRR value (higher
disorder) of s2.

Neutron diffraction data for s1 and s2 were collected at 100 K
(Fig.\ref{fig:scddata}). Data were also collected at 20 K for s2 to
examine the magnetically ordered state. The previous
crystallographic data reported by Nasch \textit{et al.}
\cite{rt2zn20struct} determined from single crystal x-ray scattering
experiments were used as starting parameters for structural
refinements. The subsequent refinements confirmed the
CeCr$_2$Al$_{20}$ type structure and produced nearly identical
atomic positions, but with small differences in the atomic site
occupancies which may indicate a slight zinc deficiency in the
samples studied here $(<1.5\%)$. The small $\chi^2$ and residual
values reported in Table \ref{tb:Feref1} indicate that these
refinements are of high quality.

Table~\ref{tb:Feref1} summarizes the refined lattice parameters,
atomic positions and site occupations of s1 and s2. No difference
was found between the lattice parameters and atomic positions of s1
and s2. Moreover, the refinements indicate that differences in the
atomic site occupations are at the level of 3$\sigma$, although both
samples appear to have a slight Zn deficiency. Hence, we are able to
conclude that any differences in site occupancy are indeed quite
subtle. Here we note that in contrast to the x-ray cross-sections
for Fe and Zn, the neutron cross-sections for Fe and Zn are differ
by a factor of $\sim$ 2.8. The anisotropic thermal parameters of s1
and s2 at 100 K are compared in table~\ref{tb:Feref2}. The
refinement results indicate universally smaller thermal parameters
for s1 than for s2. The thermal parameters are sensitive to disorder
and hence the difference between the two sets of thermal parameters
suggests that s2 is slightly less ordered than s1 consistent with
the experimental observation that s2 exhibits a lower ordering
temperature. However, given the quality of the refinements the level
of disorder is expected to be small.

s2 was measured at 20 K to examine the magnetically ordered state.
The results of structural refinements including a magnetic moment on
Tb site are shown in Table~\ref{tb:Feref1}. For consistency the
atomic fractions were restricted to the 100 K value. As at higher
temperatures the residuals and $\chi^2$ indicate high quality
refinements. An ordered moment of 6.01(5) $\mu_B$ on the Tb site was
given by the fits. Since the ordered moment would not be fully
saturated at 20 K a larger fraction of the full Tb moment should be
recovered at low temperature. We also attempted to refine a moment
on Fe in the ordered state. The refinements were not robust, but a
moment of less than 1 $\mu_B$ improved $\chi^2$ slightly. Further
experiments that can collect data at low Q values and hence are more
sensitive to the sharper fall off of the Fe form factor are required
to explicitly verify this possibility.

\section{Discussion}
In the RT$_2$Zn$_{20}$ compounds, the rare earth and transition
metal ions are surrounded by Zn cages preventing direct exchange
interactions between the 4$f$ levels or the 3$d$ levels. Rather, an
indirect exchange interaction mediated by the conduction electrons,
referred to as the Ruderman-Kittel-Kasuya-Yosida (RKKY)
\cite{RKKY1,RKKY2,RKKY3,RKKY4} exchange interaction provides the
mechanism for the magnetic order. Therefore, the number of
conduction electrons plays an important role in determining the
magnetic properties of the RT$_2$Zn$_{20}$ compounds. The remarkable
differences in magnetic behavior between TbCo$_2$Zn$_{20}$ and
TbFe$_2$Zn$_{20}$ can be associated with the fact that there are two
extra electrons per formula unit in TbCo$_2$Zn$_{20}$ than in
TbFe$_2$Zn$_{20}$. Our neutron diffraction data show explicitly that
the magnetic interactions have changed profoundly going from the
ferromagnetic interactions in TbFe$_2$Zn$_{20}$ to antiferromagnetic
interactions in TbCo$_2$Zn$_{20}$. This is consistent with the band
structure calculations \cite{nature-physics-2007}, which indicate
higher $d$-electron density of states at the Fermi level for the
RFe$_2$Zn$_{20}$ compounds than the RCo$_2$Zn$_{20}$ analogues that
result in a FM ground state for the RFe$_2$Zn$_{20}$ compounds and
an AFM ground state for the RCo$_2$Zn$_{20}$ compounds.

TbCo$_2$Zn$_{20}$ orders at $T_N$ $\approx$ 2.5 K which is as
expected due to the very low rare earth concentration and the large
distance between R-R ions in the RT$_2$Zn$_{20}$ series, the
shortest R-R spacing is $\sim$6$\AA$. However, local moment bearing
RFe$_2$Zn$_{20}$ compounds exhibit high ordering temperatures ($T_C$
$\approx$ 86, 66, 46 K for R = Gd, Tb, and Dy respectively). It has
been proposed that the 3$d$ electrons from Fe sites act as important
mediators for the R-R interaction in RFe$_2$Zn$_{20}$ system
regardless of the details of the mechanism involved in this
interaction \cite{RKKY3,RKKY4}, and the interaction between Fe 3$d$
electrons enhances the magnetic interaction between R$^{3+}$ local
moments, resulting in the remarkably high $T_C$ for
RFe$_2$Zn$_{20}$. This scenario is supported by the band structure
calculation results which predicts $\sim$ 0.67 $\mu_B$ induced Fe
moment in the ground state of GdFe$_2$Zn$_{20}$
\cite{nature-physics-2007,Jia-2008}. In particular, high sensitivity
of $T_C$ to the small sample-to-sample variation has been observed
in RFe$_2$Zn$_{20}$. Careful structural refinements of the two
TbFe$_2$Zn$_{20}$ samples s1 and s2 show no obvious site disorders,
but do show small differences in thermal parameters for the two
samples studied here. The refinement results suggest that s2 is
slightly more disordered than s1 as indicated by the larger thermal
parameters determined for s2. This is in good agreement with the
resistivity data which exhibits a lower RRR value for s2. It is not
typical that such small amount disorder would so drastically affect
the $T_C$. The $\sim$ 15 K difference in $T_C$ observed for s1 and
s2 may be related to the proximity of the RFe$_2$Zn$_{20}$ compounds
to the Stoner limit, and their associated greatly enhanced band
magnetism. As also demonstrated in doping studies
\cite{nature-physics-2007, Jia-2007}, the enhanced magnetism is
extremely sensitive to disorder so that even very small amounts of
disorder could significantly change the Curie temperature.\\
\\
\textbf{Acknowledgments}
\\
\\
We would like to thank V. O. Garlea for helping with the
TbCo$_2$Zn$_{20}$ magnetic structure refinement. Ames Laboratory is
operated for the U.S. Department of Energy by Iowa State University
under Contract No. DE-AC02-07CH111358. Work at ORNL was supported by
the Scientific User Facility Division, office of Basic Energy
Science, DOE. Work at Argonne National Laboratory was supported by
the U.S. Department of Energy, Office of Science, Office of Basic
Energy Sciences, under contract DE-AC02-06CH11357.

\end{document}